\documentclass[aps,prb,twocolumn,reprint,showpacs]{revtex4-1}
\usepackage{subfigure}
\usepackage{amsfonts}
\usepackage{amsmath}
\usepackage{amssymb}
\usepackage{graphicx}
\usepackage{color}
\begin{document}

\title{Tailoring optical fields emitted by nanometric sources.}

\author{Ra\'{u}l A. Bustos-Mar\'{u}n$^{1,2}$, Axel D. Dente$^{1}$, 
Eduardo A. Coronado$^{2}$, and Horacio M. Pastawski$^{1}$}

\begin{abstract}
In this work we study a simple way of controlling the emitted fields of sub-wavelength nanometric sources. The system studied consists of arrays of nanoparticles (NPs) embedded in optical active media.
The key concept is the careful tuning of NP's damping factors, which changes the eigenmode's decay rates of the whole array. This inevitably leads, at long time, to a locking of relative phases and frequencies of individual localized-surfaces-plasmons (LSPs) and, thus, controlls the emitted field.
The amplitude of the LSP's oscillations can be kept constant by embedding the system in optical active media.
In the case of full loss compensation, this implies that, not only the relative phases, but also the amplitudes of the LSPs remain fixed, leading us, additionally, to interpret the process as a new example of synchronization.
The proposed approach can be used as a general way of controlling and designing the electromagnetic fields emitted by nanometric sources, which can find applications in optoelectronic, nanoscale lithography and probing microscopy.
\end{abstract}

\affiliation{$^{1}$IFEG and FAMAF, UNC, Ciudad Universitaria, 5000
C\'{o}rdoba, Argentina.} 
\affiliation{$^{2}$Departamento de
Fisicoqu\'{\i}mica, Facultad Ciencias Qu\'{\i}micas, UNC, Ciudad
Universitaria, 5000 C\'{o}rdoba, Argentina.}

\maketitle

\maketitle

\section{INTRODUCTION.}

Recent advances in past decades in fabrication and characterization of nanometric devices have
given rise to a revolution, fueled by the new and intriguing properties of matter in this size scale.
Among the new fields that rapidly became central, emerged the promise of plasmonics with applications
that go from ultra sensitive nano-sensors to plasmonic circuitry.
\cite{Maier-book,Novotny-book,NanoscaleEdu,CRNordlander,SPCircuitry}
Many of those promises have became a reality nowadays, but the advances do not
seem to slow down and new ideas are still emerging in this field. One interesting example, is the
combination of plasmonic devices with active media that compensate in part or totally system's losses.
\cite{ExamGain1,ExamGain2,ExamGain3,ExamGain4,ExamGain5,ExamGain6,ExamGain7,Li,Spaser2003,Spaser2010,Spaser2011,Soukoulis,expSpasers1,expSpasers2,expSpasers3,expSpasers4,expSpasers5,expSpasers6,Spa-Mode-Sel1,Spa-Mode-Sel2}

Active media are made of dye molecules or semiconductors nanocrystals, where the population inversion is
created optically or electrically.
The concept of spaser (surface plasmon amplification by stimulated emission of radiation),
also known as surface plasmon laser in a wider context,
is an example of that. Originally proposed by Bergman and Stockman in 2003,\cite{Spaser2003} and finally implemented experimentally in 2009,\cite{expSpasers1,expSpasers2,expSpasers3} it is basically a source of electromagnetic fields, containing both propagating and evanescent waves, and formed by the interaction of surface plasmons with active media
that fully compensate the losses of the plasmonics system.\cite{Spaser2003,Spaser2010,Spaser2011,Soukoulis}

Spasers can provide us with many possibilities for prospective applications in nanoscience and nanotechnology, in particular for near-field nonlinear-optical probing and nanomodification.
In this respect, it should be desirable to control and to design \textit{a priori} the electromagnetic fields generated by those hybrid systems. If the plasmonic system consist of arrays of NPs, the design of electromagnetic fields implies a control over the synchronized oscillation of the individual localized surface plasmons (LSPs), which leads to another interesting aspect. Essentially, as in those systems not only the phases and frequencies of individual LSPs but also their amplitudes remain fixed, the whole phenomenon can be interpreted  as another example of synchronization

The phenomenon of synchronization, usually defined as the adjustment of rhythms of self-sustained
oscillating objects because of their mutual interaction,\cite{SyncBook} has been observed in
many physical and biological systems:
from the coupled pendulums clocks first described by Christian Huygens\cite{Huygens}
to the chemical or biological examples, such as fireflies that flash
in unison.\cite{SyncBook} However, up to our knowledge, it has never been described in the context of plasmonics.

In this work we study plasmonic systems, consisting of metallic nanoparticle (NP) arrays, where losses are partially or fully compensated by an active medium.
We not only find that the localized surface plasmons (LSPs) of individual NPs can be kept oscillating
with a fixed amplitude and a fixed relative phase, becoming a new example of synchronization,
but we also show that it should be relatively easy to control their asymptotic states by controlling NP's damping.
The manipulation of the system's state at long time implies the control of NP's dipolar moments and thus of their emitted electromagnetic field.
Therefore, our approach is a general way of designing the interference patterns of sources of optical fields in the sub-wavelength scale, which can have applications in several areas of nantechnology.

The paper is organized as follows: In section \ref{SecCDA} we develop the basics tools
used in our calculation. In section \ref{SecResults} we present the main results, analyzed through two simple examples of NP's arrays, and discuss them in terms of: non-Hermiticity of the dynamical matrix and asymptotic states, phase and frequency locking, role of active media, gain-loss compensation, amplitude locking, and generalization to more complex structures, subsections \ref{subSecR0} to \ref{subSecR5}. Finally, in section \ref{SecConclusions} we summarize the main conclusions.

\section{COUPLED DIPOLE APPROXIMATION FOR ELLIPSOIDS WITH RADIATION DAMPING.}
\label{SecCDA}
The systems studied are basically different arrays of metallic NPs which are modeled through the
well known coupled dipole approximation.\cite{Bustos1,Bustos2,CDA1,CDA2,CDA3,CDA4,CDA5}
In this model, each $i^{\mathrm{th}}$-NP is described by a dipole $\vec{P}_{i}$
induced by the electric field produced by the others dipoles, $\vec{E}_{j,i}$, and the external source,
$\vec{E}_{i}^{(\mathrm{ext})}$. We assume a generic ellipsoidal shape for the NPs whose
polarizabilities $\alpha$ are described in a quasi-static approximation,
\cite{Ellipsoids,KellyCoronadoSchatz}
\begin{equation}
\alpha=\frac{\epsilon _{0}V(\epsilon -\epsilon _{m})}{\left[
\epsilon _{m}+L(\epsilon -\epsilon _{m})\right] } \label{alpha},
\end{equation}
where $V$ is the volume,
$\epsilon_{0}$ is the free space permittivity, $\epsilon_{m}$ is the dielectric constant
of the host medium, and $L$ is a geometric factor that depends on the shape of the ellipsoidal
NP and the direction of $E$. The dielectric constant of the NP, $\epsilon$, is described by a Drude-Sommerfeld's like model
\begin{equation}
\epsilon = \epsilon_{\infty}-\frac{\omega_{_\mathrm{P}}^{2}}{(\omega ^{2}+i\omega \eta )},
\end{equation}
where $\epsilon_{\infty}$ is a material dependent constant and take into account 
the contribution of the bound electrons to the polarizability, $\omega_{_\mathrm{P}}^{}$ is the
plasmon frequency, and $\eta$ the electronic damping factor. Assuming for simplicity a linear array of NPs and a near
field approximation for $\vec{E}_{i,j}$ yields,
\begin{equation}
\vec{E}_{i,j}=- \frac{\gamma ^{T,L} \vec{P}_{j}}{4\pi \epsilon _{0}\epsilon_{m}d^{3}}, \label{E-quasi}
\end{equation}
where  $d$ is the distance between NPs, and $\gamma$ is a constant that
depends on the orientation of the NP's array relative to the direction of $E$,
$\gamma ^{T}=1$ if it is perpendicular and $\gamma ^{L}=-2$ if it is parallel.
If we take take into account these considerations, then all $\vec{P}$s and $\vec{E}^{(\mathrm{ext})}$s can be arranged
as vectors $\mathbf{P}$ and $\mathbf{E}$ resulting in:\cite{Bustos1,Bustos2}
\begin{equation}
\mathbf{P} = \left( \mathbb{I}\omega ^{2}-\mathbb{M}\right)^{-1} \mathbb{R} \mathbf{E}=
\mathbf{\chi} \mathbf{E},\label{MatrixP}
\end{equation}
where $\mathbf{\chi}$ is the response function, $\mathbb{M}$ is the dynamical matrix and
$\mathbb{R}$ is a diagonal matrix that rescales the external applied field according to local
properties:
\begin{equation}
R_{i,i}=-\epsilon _{0}V_{i} \omega_{_\mathrm{P}i}^{2} f,\label{Rii}
\end{equation}
with 
\begin{equation}
f= \frac{ \left[ 1 - 
(\epsilon_{\infty}-\epsilon _{m,i}) \left( \omega ^{2}+i\omega \eta_{i}\right)
/\omega_{_\mathrm{P}i}^{2} \right] }
{\left[\epsilon _{m,i}+L_{i}(\epsilon_{\infty}-\epsilon _{m,i})\right]}. 
\end{equation}
To understand the physical meaning of $f$, first note that Eq. \ref{MatrixP} resembles that of a set of coupled harmonic oscillators. In the quasi-electrostatic limit, Eq. \ref{E-quasi}, and for a negligible radiation damping term, see Eq. \ref{Gamma}, this similarity is strict for $f$ equal to 1. Thus, this factor essentially accounts for deviations of the ideal model of coupled harmonic oscillators.

The coupling constants, $M_{i,j}= - \omega_{_\mathrm{X}i,j}^{2}$ (for $i \neq j$),
and the LSP complex square frequencies,
$M_{i,i}=\omega_{_\mathrm{SP}i}^{2}-\mathrm{i} \Gamma_i(\omega)$, are given by:\cite{Bustos1,Bustos2}
\begin{equation}
\omega_{_\mathrm{X}i,j}^{2} =\frac{\gamma ^{_{T,L}} V_{i} \omega_{_\mathrm{P}i}^{2}}
{4\pi \epsilon_{m}d_{i,j}^{3}} f,
\label{OmegaX}
\end{equation}
\begin{equation}
\omega_{_\mathrm{SP}i}^{2} =\frac{\omega_{_\mathrm{P}i}^{2}L_{i}}{\left[
\epsilon _{m,i}+L_{i}(\epsilon_{\infty}-\epsilon _{m,i})\right] },
\label{OmegaSP}
\end{equation}
and
\begin{equation}
\Gamma(\omega)=\eta \omega + \eta _R \omega ^3. \label{Gamma}
\end{equation}
where $\eta$ is the electronic damping and $\eta _R$ the radiation damping. The electronic damping $\eta$
can be calculated from the Fermi velocity $v_f$, the bulk mean free path $l_{bulk}$, the volume $V$, and the surface $S$ of the NP by using the Matthiessen's rule $\eta= v_f (1/l_{bulk}-C/l_{eff})$ with $C \approx 1$, and the
Coronado-Schatz formula $l_{eff}=4V/S$.\cite{CoronadoSchatz}
The value of $\eta _R$ can be calculated from the ellipsoid's radius $a$, $b$, and $c$,
$\eta _R=2/9(a b c/v^3) \omega_P^2 f$, where $v$ is the speed of light in the host medium.
This extra damping term appears when the polarizability $\alpha$ is corrected by using the modified
long-wavelength approximation, $\alpha'=\alpha [1- i (2/12 \pi \epsilon_0) k^3 \alpha]^{-1}$.\cite{KellyCoronadoSchatz}
In the examples analyzed here, dynamic depolarization is negligible and thus not included
in the equations for simplicity.

Retardation effects change the coupling terms which now should be determined by the
true dipole-induced electric field, \textit{i.e.}:
\begin{align}
\vec {E} =  \frac { e^{ ik d } |P|} { 4\pi \epsilon_0 \epsilon_m d^{ 3 } }
& \left\{ (k d)^{ 2 }(\hat { d } \times \hat { p } )\times \hat { d }   \right. \\
& + \left. \left[ 3\hat { d } (\hat { d } \cdot \hat { p } )-\hat { p }  \right] \left( 1- ikd \right)  \right\},
\end{align}
where $k$ is the wavenumber in the dielectric, $k=\omega/v$ (where $v$ is the speed of light in the medium),
$\hat { d }$ is the unit vector in the direction of $\vec {d}$ (where $\vec {d}$ is the position of the
observation point with respect to the position of the dipole), $\hat { p }$ is the unit vector in the direction of $\vec {P}$, and $|P|$ is its modulus.
If the system consists of a linear array of NPs where the spheroids axes are aligned with respect to the direction of the array, transversal ($T$)
and longitudinal ($L$) excitations do not mix, which allows us to preserve the form of Eq. \ref{OmegaX}
by simply replacing $\gamma^{_{T,L}}$ by $\widetilde{\gamma }^{_{T,L}}$, where:

\begin{align}
& \widetilde{\gamma }^{_{L}}_{i,j}=-2[1-ikd_{i,j}]e^{ ikd_{i,j} } \notag \\
& \widetilde{\gamma }^{_{T}}_{i,j}=[1-ikd_{i,j}-(kd_{i,j})^{ 2 }]e^{ ikd_{i,j} }.
\end{align}

We use this final form of the equations in all the calculations shown here. However, the qualitative results do not change by using the quasistatic approximation.

The temporal evolution of the dipolar moments of individual NPs can be evaluated
by Fourier transforming the response function $\chi(\omega)$ into $\chi(t)$ and 
using the convolution theorem:
\begin{equation}
P_i(t) = \sum_{j} \int_{0}^{t} \chi_{i,j} (t-\tau) E_{j}^{(ext)}(\tau) d \tau. \label{fourier} 
\end{equation}
The functions $\chi(t)_{i,j}$ were numerically computed from $\chi(\omega)_{i,j}$
by using a fast Fourier transform algorithm.\cite{Numerical} Here, one must be careful,
in case of using active media, of not overpassing the loss-compensation condition as
one is always assuming that the response function $\chi_{i,j}$ is square integrable.

\section{RESULTS.} \label{SecResults}

\subsection{Non-Hermiticity of $\mathbb{M}$ and asymptotic states.}
\label{subSecR0}
In the type of system studied here, frequency and phase locking may appear as a natural
consequence of the properties of non-Hermitian matrices.
While isolated systems are described by a typical Hermitian dynamical matrix
$\mathbb{M}$, where the final state depends on the initial conditions, the
presence of an ``environment'' leads to a non-Hermitian dynamical matrix. \cite{Bustos1,Bustos2,Rotter,PastPhysB}
This interaction may cause asymptotic states that are independent of the initial conditions.
An illustrative example of that is the case of a pair of piano strings in a
unison group.\cite{pianos}
There, the slightly detuned strings are coupled through the bridge,
which, in turn, is coupled to a dissipative soundboard.
Within a certain critical parametric range, this dissipative coupling induces the
synchronous oscillation of both strings \cite{pianos} and gives the piano
its characteristic and persistent aftersound.
This dissipative coupling can be modeled by an imaginary coupling which, at a critical
strength, produces the collapse of the pair of originally mistuned eigenfrequencies into a single tone. Simultaneously, the  originally identical dampings split into a short and a long lived modes.
The effect of this is that the long time evolution is dominated, for almost any initial condition,
by the normal mode whose eigenvalue has the smallest imaginary part.

The same analysis can be straightforwardly applied to plasmonics systems represented
by Eq. \ref{MatrixP}, where the analogy also includes the concepts of dissipative couplings, frequency collapses, and damping's splittings, see Appendix.
However, in plasmonic's systems it is not always obvious which is the asymptotic state of a given system, which makes its control even less obvious.
The situation worses if we consider that usually parameters such as NP's shape and separations are not accurately determined.
Besides, unlike the discussed case of coupled piano strings, the amplitude of the
oscillations of the LSPs decays so fast that it would be quite difficult to observe
the phase and frequency locking.
Therefore, two main features are desirable. The control at will of the asymptotic state of the system and to be able to keep
the amplitude of the LSP's oscillations over a long period of time.
In the following sections we will address sequentially each one of these points.

\subsection{Phase and frequency locking.}
\label{subSecR1}
\begin{figure}[ht]
\includegraphics[width=3.5in]{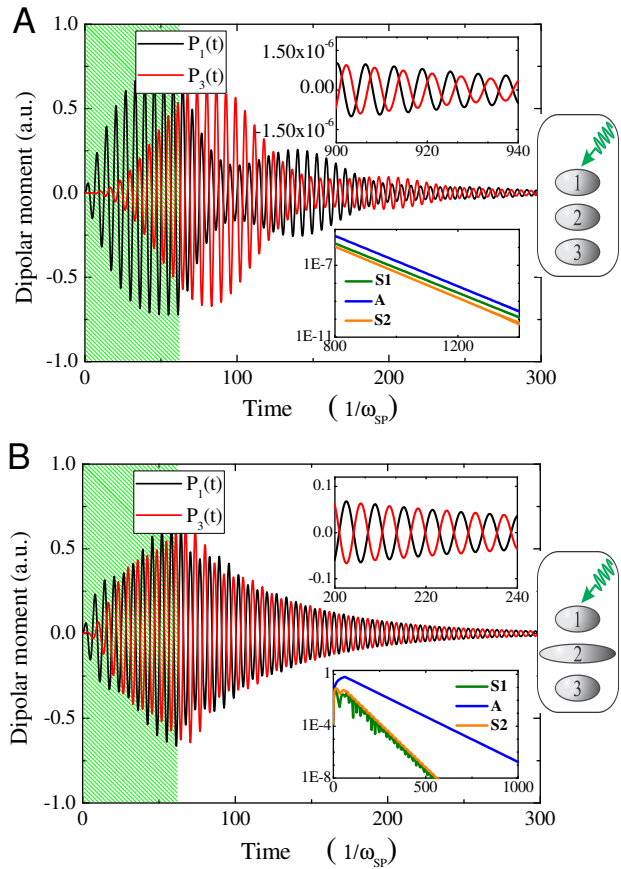} 
\caption{(Color online) \textbf{- A)}. Dipolar moment $P_i$ of NPs 1 and 3 (in arbitrary units)
vs time (in units of $\omega_{SP}^{-1}$) for three aligned and identical NPs.
Between $t=0$ and $62$ (mark in green) a external field of frequency $\omega=\omega_{SP}$
and direction parallel to the array, is applied locally to the first NP to initialize the system.
The parameters used correspond to spheroidal Ag's NPs of radii $30$, $30$ and $8$ nm,
separated $32 nm$, and $\epsilon_m=1.77$.
\textbf{B)}. The same but with the middle NP
having a different shape ($90$x$90$x$8$ $nm$).
\textbf{Upper insets}: Detail of the main figure.
\textbf{Bottom insets}: Detail of the decay rate of different oscillation modes.
$S1=(P_1+\sqrt{2}P_2+P_3)/2$, $A=(P_1-P_3)/\sqrt{2}$, and $S2=(P_1-\sqrt{2}P_2+P_3)/2$.
\textbf{Side figures}: Schemes of the NP's arrays.}
\label{Figure1}
\end{figure}

As we mentioned, the plasmonic dynamical matrix $\mathbb{M}$ resembles that of coupled harmonic oscillators. This can be used to analyze certain systems in simple terms as we will see. Assuming the quasi-electrostatic limit, negligible damping terms, and $f \approx 1$, it is easy to evaluate the normal modes of $\mathbb{M}$. In the case of three equal NPs, aligned linearly, and equally spaced the normal modes can be written as: $(P_1-P_3)/\sqrt{2}$, $(P_1+\sqrt{2}P_2+P_3)/2$ and $(P_1-\sqrt{2}P_2+P_3)/2$. Where $P1$, $P2$, and $P3$ stand for the dipolar moments in some given direction of NPs 1, 2, and 3 respectively.
If the difference in frequency between the NPs of the ends and the central one is small this expressions are still approximately valid.

Let us analyze this simple example of three aligned Nps and let us assume that we want to ensure an asymptotic state in which the NPs of the ends remain oscillating in anti-phase.
In this case, one only need to add a larger damping factor to the middle NP.
The normal mode of $\mathbb{M}$ that has zero weight over the NP with a high damping factor, $\approx (P_1-P_3)/\sqrt{2}$, has a small decay rate compared with the other two, $\approx(P_1+\sqrt{2}P_2+P_3)/2$ and $\approx(P_1-\sqrt{2}P_2+P_3)/2$, which both have finite weights over the highly dispersive nanostructrure (NP 2 in this example).
The strategy is then clear, the key to control the phase and frequency locking is the careful designing of the damping factors of NPs in such a way that it leaves one normal mode (the one that will define the desirable phase relationship and frequency) with the smallest, ideally zero, weight over the regions of the array with the largest damping factors.

There are of course several ways of increasing the damping factor
of NPs, not only by changing their shape or material but also by ``connecting'' them to waveguides for example.\cite{Bustos1,Bustos2}
Here we use the shape of NPs to control the damping factors. According to the parameters chosen, the
radiation damping term is the dominant one for the NP with the high damping factor, while the electronic damping
term is the dominant one for the others.

In Fig. \ref{Figure1}, we evaluate the temporal evolution of the dipolar moment 
$P_i(t)$ of each NP, by using Eq. \ref{fourier} in two examples that illustrate how
tuning the damping factor of NPs can be used to control the asymptotic state of
the system. In these examples we explicitly take into account the material
and shape of NPs, always within the couple dipole approximation described in section
\ref{SecCDA} and including the full dependence of $\omega_{_\mathrm{X}}^{2}$ and
$\Gamma$ on $\omega$.
The results essentially show the above discussed: After the external source is
switched off the LSPs decay very fast, but as indicated in the lower insets, different
modes decay at different rates which leads to a natural phase and frequency locking of the LSPs of 
individual NPs. 
The asymptotic state of case \textbf{A} is not easily seen in the upper inset, but a more
careful analysis, depicted in the lower inset, reveals that the mode with the lowest decay rate
is ``S1''. A comparison of Figs. \textbf{A} and \textbf{B} shows that the asymptotic state changes
as consequence of the increased damping factor of the middle NP.
It should be mentioned that the only role of the external source of electric field is just to initialize the system. This could have been done in many different ways in the simulation, for example by using a pulse of electromagnetic radiation. However, as long as all normal modes are excited the final state will be the same, up to a factor in the amplitude of course.

\subsection{Role of active media.}
\label{subSecR2}
As previously mentioned, there is a problem with the phase and frequency locking mechanism described above.
Everything occurs too fast.
Note that in Fig. \ref{Figure1} the time scale is in units of $\omega_{SP}$ of NPs 1 which for the NPs used
corresponds to around 0.2 fs.
This implies that all the process starts and finishes in less than 0.1 ps approximately.
There is a need, then, of keeping the system oscillating for longer periods of time, in order to reasonably envision possible applications.
This can be done by embedding the system in an optically active medium.

If the gain of the active medium is below the loss compensation threshold, its effect can be
modeled phenomenologically on the basis of classical electrodynamics without taking into account
explicitly the quantum dynamics of the chromophores. This is done by considering the medium as a
dielectric with a negative imaginary part in the refraction index $n$.
$n=n_0-i \kappa$.\cite{ExamGain1,ExamGain2,ExamGain3,ExamGain4,ExamGain5,ExamGain6,ExamGain7,Li,Spaser2010}
Within this model, the active medium is consistent with an homogeneous distribution of the dye molecules, or nanocrystals quantum dots, and with a wide band approximation for its response.

The wide band approximation implies that the eigenfrequencies of the modes are close compared with the frequency dependence of the active medium. If this condition is not fulfilled, each mode will have a different value of $\kappa$ or even a null one if the frequency of the mode is far enough from the maximum of the medium's stimulated-emission-spectrum. In this case the analysis of mode's compensation is direct as it can be based only on mode's frequencies. On the contrary if the eigenfrequencies of the modes are close enough, such as all modes experience approximately the same value of $\kappa$, it is in principle not obvious which mode will be compensated first, and less obvious how to control this. This is why, the wide band approximation allows us to explore alternatives for controlling the system's asymptotic states, beyond the mechanisms based on the frequency response of the active medium or the use of some spatial inhomogeneities in its distribution around the system.\cite{Spa-Mode-Sel1,Spa-Mode-
Sel2}

As mentioned, it could result not obvious how an active medium would affect the phenomenon depicted in Fig. \ref{Figure1}, mainly because $n$ enters not-linearly in the equations,
see Eqs. \ref{MatrixP}-\ref{Gamma}, and this could in principle changes the expected asymptotic state.
However, as we are precisely considering gain media without explicit spacial distribution or frequency dependence, it is reasonable to expect that all modes will be excited similarly. Thus, if there are appreciable differences in the natural decay rates, the asymptotic states with active medium should be determined directly by them.

\begin{figure}[ht]
\includegraphics[width=3.5in]{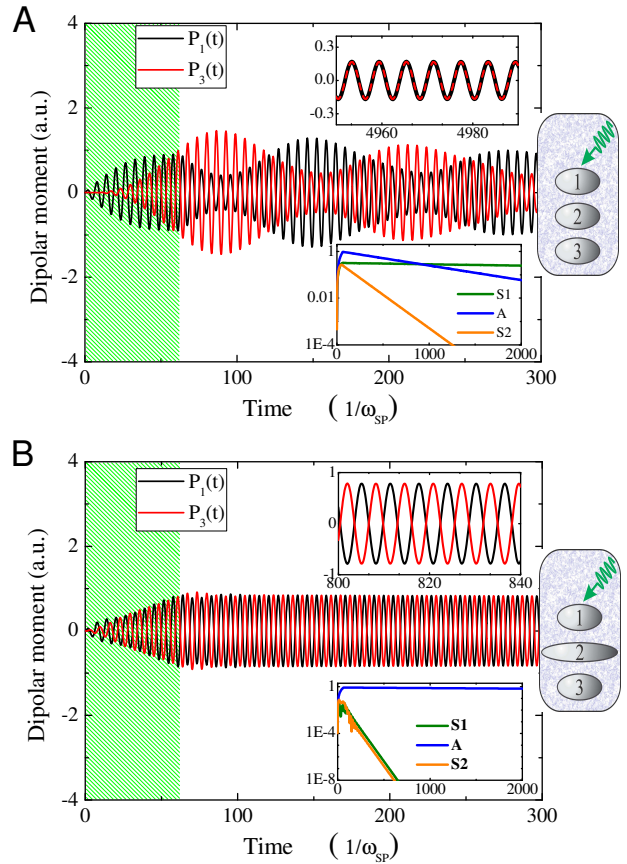}
\caption{(Color online) - The same as Fig. \ref{Figure1} but considering an optically active
medium with $\kappa=0.11$ and $0.12$ for subfigures \textbf{A} and \textbf{B} respectively.}
\label{Figure2}
\end{figure}

Figs. \ref{Figure2} shows essentially that. Incorporation of optical gain media does not change
the asymptotic states discussed in the previous section, even thought it has been used a value of $\kappa$ that almost completely compensate losses.
In the two examples analyzed, the slowest decaying mode keeps as such, modes ``S1''
and ``A'' for cases \textbf{A}, \textbf{B} respectively.
The only effect of the active medium in those examples, besides keeping the system oscillating for longer periods of times, is that it systematically increases even further the differences in the decaying rates, making phase and frequency locking to occur even earlier.
As the system remains oscillating for longer periods, it is easier to see in the figures (upper insets) the phase locking and how it is affected by changing the damping factors.
In case \textbf{A}, the NPs of the edges (Nps 1 and 3 in Fig. \ref{Figure2}) end oscillating in phase, while, if we increase the damping factor of the middle NP as in case \textbf{C}, the NPs of the edges end oscillating in anti-phase.
As mentioned before, the reason of that is simply that the anti-phase oscillation of the NPs of
the edges interfere destructively over the NP of the middle where the largest damping factor
is present. The other two normal modes having some weight on the middle NP will increase their decay rate.

Besides the examples shown in the figures, we also tried other possibilities as, other NP's arrays or changing the system's parameters.
However, the results were always the same, when there are appreciable differences in the decay rates with $\kappa = 0$, which is for example the case of Fig. \ref{Figure1}-\textbf{B}, an homogeneous active medium is not able to change the expected asymptotic state.
Only, for systems like case \textbf{A} of Fig. \ref{Figure1}, where the decay rates for $\kappa = 0$ are very close, we observed, for some system's parameters, that the active media changes the expected asymptotic state.

\subsection{Gain-loss compensation.}
\label{subSecR3}

At this point, it is important to discuss about the limiting value of $\kappa$, $\kappa_{\mathrm{lim}}$,
for which losses are exactly compensated, and the experimental feasibility of
this. The value of $\kappa_{\mathrm{lim}}$ can be evaluated from the poles of Eq. \ref{MatrixP}
by looking for the pole with the smallest imaginary part. Thus, $\kappa_{\mathrm{lim}}$ is the value of
$\kappa$ for which the imaginary part of this pole equals zero.
In some cases, it can be easy to obtain approximate analytical expressions, but in general one must resort to 
numerical evaluations.

In case \textbf{B} of Figs. \ref{Figure1} and \ref{Figure2}, the eigenvalue of the ``A``
eigenmode $\omega_{\mathrm{eig-A}}^2$, $(P_1-P_3)/\sqrt{2}$, can be obtained easily by assuming
a wide band approximation:
\begin{equation}
\omega_{\mathrm{eig-A }}^2 \approx \omega _{_\mathrm{SP}}^{2}-\mathrm{i} \Gamma,
\label{polos}
\end{equation}
where $\omega _{_\mathrm{SP}}^{2}$ is the LSP resonant frequency of one of the
NPs of the ends, and $\Gamma$ is its damping factor.
Then, the value of $\kappa_{\mathrm{lim}}$ can be obtained by using Eq. \ref{OmegaSP} with
$\epsilon_m=n^2$ and assuming a small $\kappa$. The result is:
\begin{equation}
\kappa_{\mathrm{lim}} \approx \frac {\Gamma [n_0^2+L(\epsilon_{\infty}-n_0^2)]}
{2 n_0 \omega_{_\mathrm{SP}}^2 (1-L)} \label{klim}
\end{equation}
which according to the parameters used, $n_0=1.33$, $L=0.689$, $\epsilon_{\infty}=3.7$,
and $\Gamma /  \omega_{_\mathrm{SP}}^2  \approx 0.032$, gives $\kappa_{\mathrm{lim}} \approx 0.121$.

For case \textbf{A} of Figs. \ref{Figure1} and \ref{Figure2},
it is more difficult to obtain simple analytical solutions as $ \omega_{_\mathrm{X}}^2$
also enters into the equations and depends on $\kappa$. However, they can always be evaluated
numerically. From the simulation, we estimated the value of $\kappa_{\mathrm{lim}}$ as
$0.11$ and $0.12$ approximately for cases \textbf{A} and \textbf{B} respectively, which should be close to experimental possibilities.\cite{expSpasers1,expSpasers2,expSpasers3,expSpasers4,expSpasers5,expSpasers6,GainExp1,GainExp2,GainExp3,GainExp4} Note the agreement between the numerical and analytical results for case \textbf{B}.

The value of $\kappa$ is a phenomenological coefficient that represent the property
of some media of coherently amplify an electromagnetic field. It is related with the amplification coefficient $g$ by $g = 4 \pi \kappa / \lambda$.
Gain media in plasmonics are made of chromophores that overlap spatially and spectrally with
the surface plasmon modes of the nanostructure.
These chromophores can be semiconductors nanocrystals, dye molecules, rare-earth
ions, or electron-hole excitations of a bulk semiconductor.
The gain coefficient can be written as $g = N \sigma_e$, where $N$ is the concentration of electron-hole pairs in the case of semiconductors or the concentration of molecules and their population inversion in the case of dye molecules. The symbol $\sigma_e$ is the stimulated emission cross section which, in turn, depends on the dipolar moment of the transition.\cite{ExamGain1,ExamGain2,ExamGain3,ExamGain4,ExamGain5,ExamGain6,ExamGain7,Li,Spaser2003,Spaser2010,Spaser2011,Soukoulis}

Here we should clarify one point. Up to now, we have been discussing and comparing the decay rates of different modes that have always the same direction of the electric field, parallel to the array. However, there are two other set of modes, those with the electric field perpendicular to the array, that can also enter in the analysis of the system's asymptotic state.
If the dipolar moment of the transition of the molecules or semiconductors nanocrystals that constitute
the active media, have a preferential direction, then the media can only feedback some modes, those
with a finite overlap between the mode's electric field and the dipolar moment.\cite{Spaser2010,Spaser2011,Soukoulis} In this case, only some modes, ideally those that oscillate in the preferable direction, should be considered.
On the contrary, if the dipolar moment of the transition has a random orientation,
then one has to analyze the full picture, i.e. the whole nine modes for the arrays.
In this last case, the shape of the NPs acquires a central role, because it determines in which direction the system will remain oscillating. To see that, note that the $L_i$ factor of Eq. \ref{alpha}, depends on both the shape of the NP and the direction of electric field, and this parameter enters, not only in eigenfrequency of the mode, but also in the damping term $\Gamma$ through $\eta_R$ and $f$.

For example, let us consider a system of three equal NPs of 30x20x8 nm aligned in the direction of the minor axis and separated 24 nm. Here, the mode that is compensated in the first place by the active medium is that where all LSPs are oscillating synchronous in phase and parallel to the major axis. In this case $\kappa_{\mathrm{lim}} \approx 0.024$. The equivalent modes but for the other directions, those where the LSPs oscillate in phase and parallel to the second largest axis and to the minor axis, have a value of $\kappa_{\mathrm{lim}}$ of approximately $0.029$ and $0.073$ respectively.

\subsection{Amplitude locking.}
\label{subSecR4}
\begin{figure}[ht]
\includegraphics[width=2.7in]{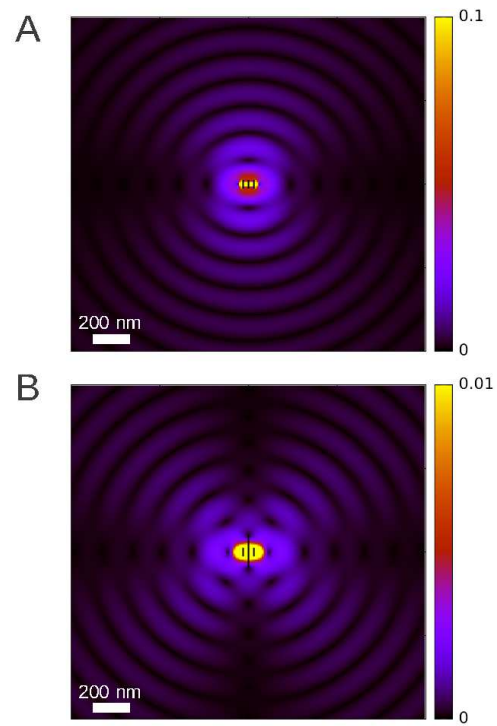}
\caption{Electric field $E$ for the asymptotic state of case \textbf{A} and \textbf{B} of Fig. \ref{Figure2}. The strength of $E$ is normalized to its maximum value in each figure.}
\label{Figure3}
\end{figure}
According to our equations up to now, for $\kappa > \kappa_{\mathrm{lim}}$, $P(t)$ should grow exponentially \textit{at infinitum}, which is of course not realistic. At some point the pumping mechanism that keeps the inversion
population must be overcome by the decay rate of the molecules in the excited state decaying
toward their fundamental state.
The realistic situation is that the amplitude of the surface plasmon oscillations should stabilize at some point. This is so, because the stimulated emission that depletes the excited states depends on $|E|^2$ which, in turn, depends on $P_i$, while the mechanism that restore the inversion population is fixed and independent of $P_i$.\cite{Spaser2010,Spaser2011,Soukoulis}

A complete treatment would require to solve the quantum mechanics dynamics of each chromophore
under the influence of the electromagnetic field corresponding to its position and the coupled
equation of motion of the surface plasmon dynamics. This is beyond the scope of this work and besides
was already addressed by other authors in the context of spasers.\cite{Spaser2010,Soukoulis} The important
result of these previous works, for the present purposes, is that the system evolves in a somehow complex way, until a stationary regime is reached.
This stationary regime corresponds to a net amplification equals zero, which means that gain exactly compensate losses,\cite{Spaser2003,Spaser2010,Spaser2011,Soukoulis} a condition expressed in our case by Eqs \ref{klim} in terms of $\kappa_{\mathrm{lim}}$.
Essentially, the convergence towards a stationary regime where losses are compensated,
implies amplitude locking. The asymptotic value of the amplitude could be complex to evaluate
but the important point is that, sooner or latter, it is reached and it is non zero for $\kappa_{\mathrm{initial}} > \kappa_{\mathrm{lim}}$ and initial conditions different from the trivial one, $P_i = 0$.
The other important point is that, once the system is in the stationary state regime, the inversion population freezes, fixating the gain coefficient $g$ and thus $\kappa$, at $\kappa = \kappa_{\mathrm{lim}}$.
Then, independently of how or when this stationary regime is reached, in the end one should see the type of behaviors showed in the context of Fig. \ref{Figure2}, i.e. different normal modes are compensated differently by the active medium.
Therefore, while the slowest decaying mode is exactly compensated, the others will be undercompensated which will inevitably lead to a phase, frequency, and also amplitude locking. Note that, because of this, the plasmonics systems studied can be considered as a new example of synchronization.

The above analysis has also another important consequence, gain medium can not, in general, exactly compensate the losses of all eigenmodes at the same time. Let us assume the system has three eigenmodes each one with different values of $\kappa_{\mathrm{lim}}$; $\kappa_{1} < \kappa_{2} < \kappa_{3}$ . Then, if one try to compensate the second or the third modes , $\kappa = \kappa_{2}$ or $\kappa = \kappa_{3}$, the first one will be overcompensated which can not define a stationary state as it should grow indefinitely. The realistic situation is that the inversion population of the active medium will be depleted by the increasing electromagnetic field of the first mode, reducing the value of $\kappa$ until it reaches $\kappa_{1}$.
As this argument is very general, we believe its consequences should be present in the majority of this kind of systems provided that the necessary ingredients are present. The eigenfrequencies of the modes must be close enough compared with the frequency response of the medium and different modes should share somehow the same dye molecules or semiconductor nanocrystals. We plan to address this interesting issue in a future work.

Figs. \ref{Figure3}\textbf{-A} and \ref{Figure3}\textbf{-B} show the electric field generated by the examples shown in Figs. \ref{Figure1} and \ref{Figure2} for $t \rightarrow \infty$. The former corresponds to the system with three equal Nps and the latter to the system with the middle Np having a larger damping factor. Note the great differences in the emitted electric fields.
The upper case shows the typical interference patterns of a punctual dipolar source, while the lower one shows that of a quadrupole.
This example highlight the fact that amplitude locking becomes our system into, not only another example of synchronization, but also a nanometric source of both evanescent and propagating waves with a predetermined and controllable interference pattern.

\subsection{Generalization to more complex structures.}
\label{subSecR5}
\begin{figure}
\includegraphics[width=3.0in]{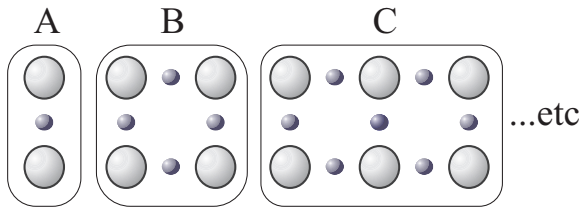}  
\caption{(Color online) - Schemes of others NP's arrays. Small spheres stand for NPs with
large damping factors while the large ones represent NPs with small damping factors.}
\label{Figure4}
\end{figure}
The proposed synchronization mechanism can be easily extended to more complex nano-structures.
The key is to build the system such as all normal modes but one have some weight on the highly
dispersive NPs, the middle one in Fig. \ref{Figure2}-\textbf{B} for example.
Then, if the damping factor of the highly dispersive NP is large enough, 
the slowest decaying normal mode, which will control the relative phases of the LSP in the asymptotic state,
will be that having the smallest weight over these NPs.
In Fig. \ref{Figure4} we present just one possible set of examples of that for NP's arrays
of arbitrary size. The examples assume nearest neighbors interactions.
The small spheres represent equal NPs with large damping factors while the large ones
represent equal NPs with small damping factors.
In all these cases, one can show that there is always one eigenvalue of $\mathbb{M}$ that has
zero weight over the small NPs. This normal mode corresponds to that where the LSP of the large NPs
oscillates in anti-phase with respect to their nearest large-NPs neighbors.
Thus, as this mode will have the slowest decay rate it will determine the phase locking at long
time. Others asymptotic states are also possible in those systems.
One only has to evaluate the weight of individual NPs on each normal mode and, based on that, increases
selectively the damping factors of certain Nps to achieve the desired asymptotic state.

\section{CONCLUSIONS.}
\label{SecConclusions}
In this work we have shown a simple way of controlling phase and frequency locking of the self-sustained oscillation of NP's LSPs, by tuning the damping factors of individual NPs.
Furthermore, we have shown that it should be possible to keep the system oscillating with constant
amplitude by including optically active media properly tuned.
We interpret this as a new example of synchronization as we are in the presence of self sustained
oscillating objects, clearly separable, that depict phase, frequency as well as amplitude
locking, consequence of their mutual interaction.
Since it is possible to control the asymptotic state of these NP arrays with self sustained
LSP, our approach is a general way of designing the interference patterns of sources of optical
fields in the sub-wavelength scale. This can surely find applications in optoelectronic, nanoscale lithography and probing microscopy.
In addition, the proposed method can naturally be combined with other alternatives, such as using the frequency dependence of the active medium or controlling its spatial distribution.

\section{ACKNOWLEDGEMENTS.}
The authors acknowledge the financial support from CONICET, SeCyT-UNC, ANPCyT,
and MinCyT-C\'{o}rdoba. E.A.Coronado thanks the financial support provided by  CONICET PIP (2012) 112-201101-00430 and by FONCYT Program BID PICT 2012-2286.

\newpage

\section{Appendix: Dissipative couplings and dynamical phase transitions.}
We mentioned that in the case of coupled piano strings, there is a 
dissipative coupling between the strings which can be modeled by an imaginary coupling term in the dynamical matrix $\mathbb{M}$. Pure imaginary, or at least complex, couplings have interesting effects on the properties of the eigenvalues of $\mathbb{M}$. At some critical values of the system's parameters, there can be a collapse of the real part of the eigenvalues of $\mathbb{M}$ and a bifurcation of their imaginary part at points called ''exceptional points``. There, among other effects, $\mathbb{M}$ becomes singular and the system's eigenvectors behave
oddly in their surroundings.\cite{Rotter,Nimrod}
Since the dynamical observables have a non-analytic dependence on the system's paremeters, this results in what is called a dynamical phase transition, DPT.\cite{Rotter,PastPhysB,Bustos1,Bustos2}  

In the case of plasmonics systems, as those showed in this work, the complex coupling
can be seen as just the consequence of the effective interaction between two parts of a system
connected through a bridging dissipative subsystem.
For example, if we have three NPs aligned one can always calculate an effective coupling between the NPs at the ends.\cite{RevMex} The result of this is a complex effective coupling, consequence of the damping term of the NP in the middle.\cite{Bustos1,Bustos2}

\begin{figure}[p,h]
\centering
\includegraphics[width=3.2in]{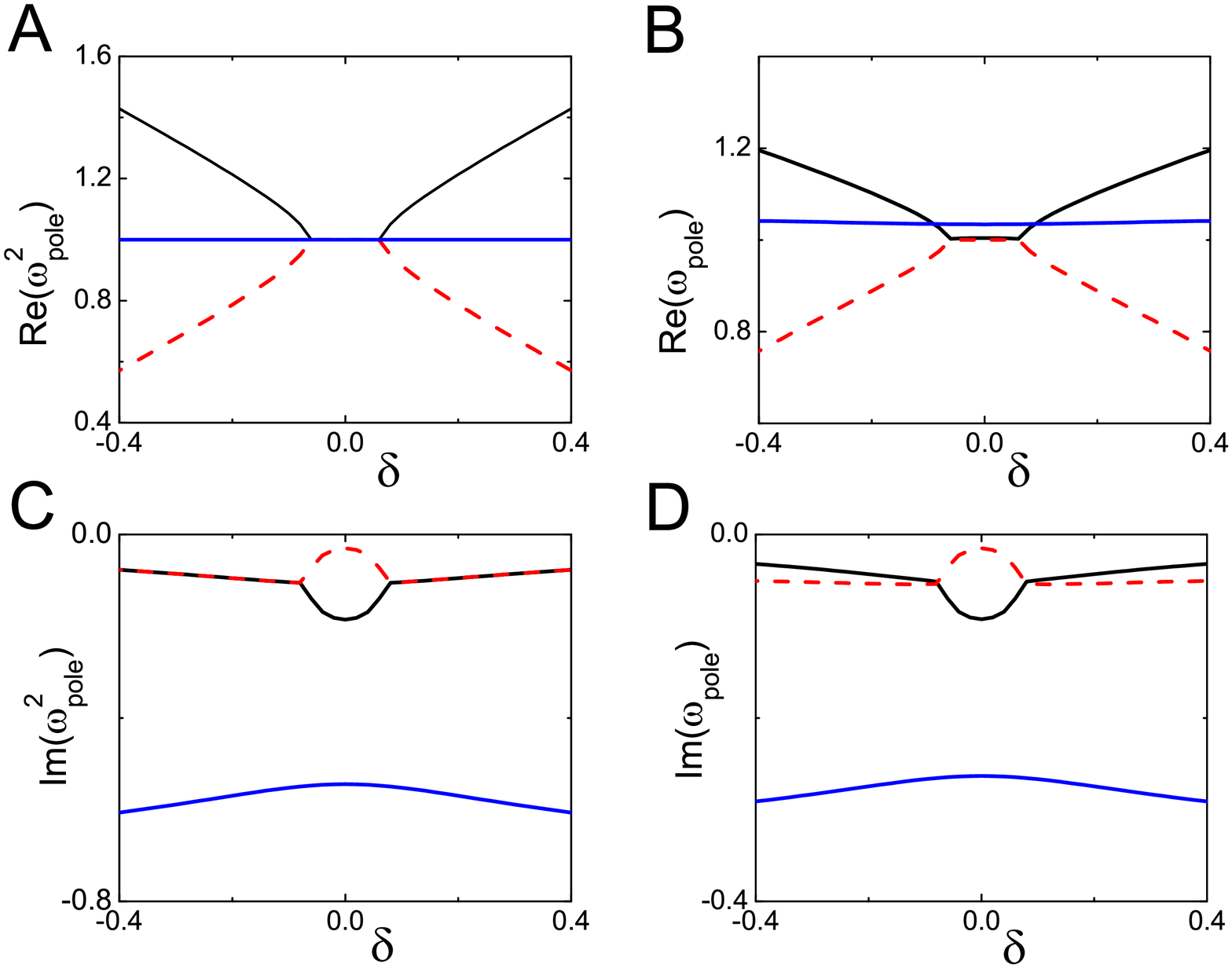}
\caption{(Color online) - Figs. \textbf{A} and \textbf{C} are respectively the real and imaginary part of 
$\omega^2_{\mathrm{pole}}$, the eigenvalues of $\mathbb{M}$; while Figs. \textbf{B} and \textbf{D} are respectively
the real and imaginary part of  $\omega_{\mathrm{pole}}$, the poles of $\chi$. The system consists of three aligned NPs with  $\delta=\omega_{_\mathrm{SP1}}^{2}-\omega_{_\mathrm{SP3}}^{3}$, where 1 and 3 stand for the NPs of the edges. Only nearest neighbor couplings are considered, $\omega_X^2 = 0.2$, $\omega_{SP}^2=1$ (for $\delta=0$), and $\Gamma=0.03$ for all NPs except for the middle one where $\Gamma=0.7$.}
\label{FigureA1}
\includegraphics[width=3.2in]{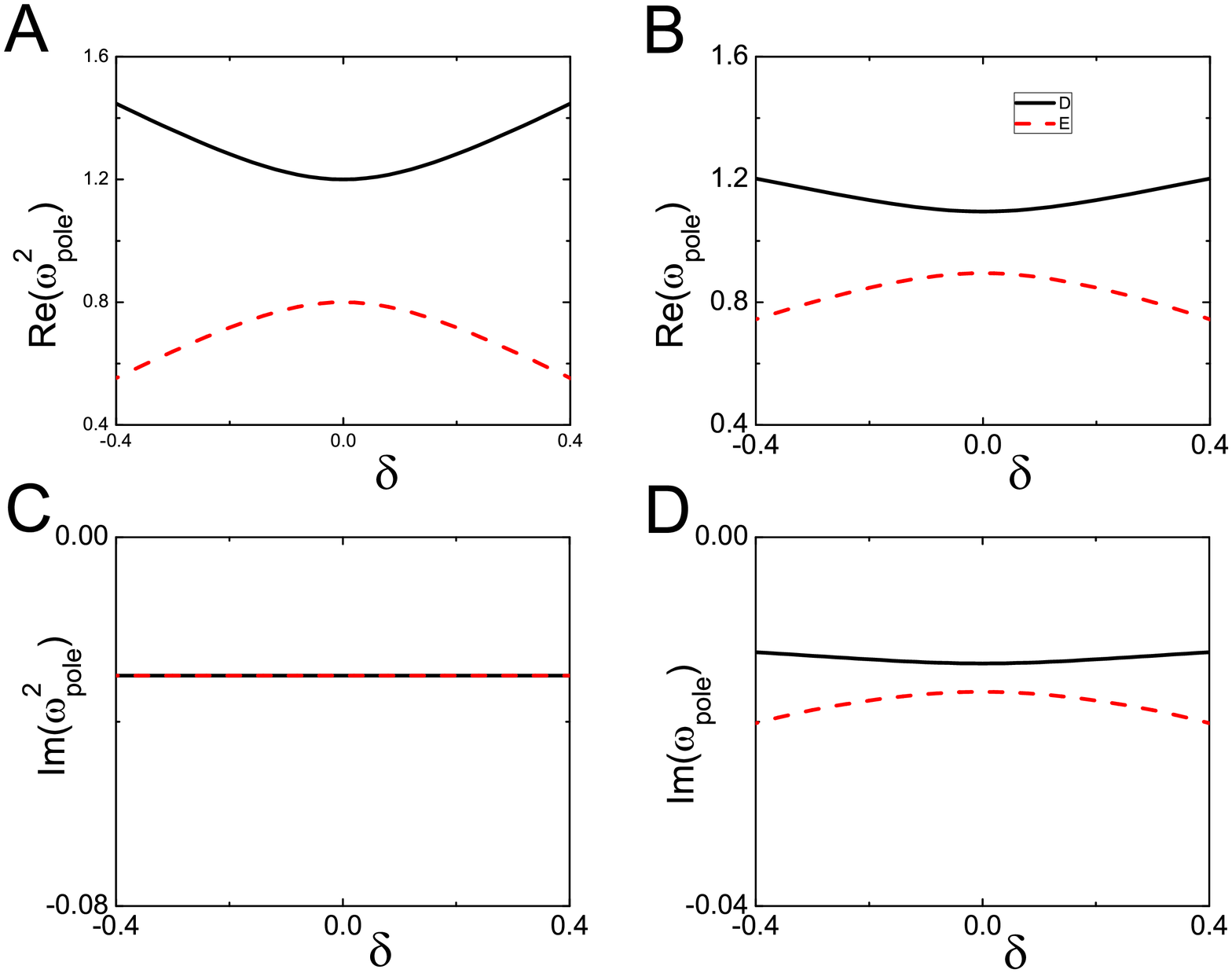}
\caption{(Color online) - The same as Fig. \ref{FigureA1} but for a system of 
two NPs with equal damping, $\Gamma=0.03$. Notice that in spite of the small differences in decay rates as compared to those in Fig. \ref{FigureA1}, they can be enough to produce an observable phase locking through the use of an active medium.}
\label{FigureA2}
\end{figure}

Fig. \ref{FigureA1} shows that  the eigenvalues of $\mathbb{M}$ present a collapse of their real part accompanied by a splitting of their imaginary part. Just as in the example of the coupled piano strings.
This case corresponds to a very large value of the damping term of the middle NP and a mistuning parameter, $\delta$, below a critical value.
Here, it should be mentioned that what really sets the decay rates, are the imaginary part of the poles,
$\mathrm{Im} \left ( \omega _{\mathrm{pole}} \right )$, of the response function $\chi(\omega )$, and
not the imaginary part of  the eigenvalues of $\mathbb{M}$. In the wide band approximation, these last coincides with $\mathrm{Im} \left( \omega_{\mathrm{pole}}^{2} \right)$.
This distinction can be quite irrelevant in some situations but becomes fundamental in others.
In  Fig.\ref{FigureA2} we consider the case of two internacting NPs. We can see that although the eigenvalues of $\mathbb{M}$ have exactly the same imaginary part, which would preclude the synchronization mechanism depicted in the main section of the article, there is a difference in the imaginary part of $\omega_{\mathrm{pole}}$.
Although this difference is very small, as compared with the case shown in Fig. \ref{FigureA1},
it is enough to give rise to a characteristic asymptotic state and, thus, it can be used to induce a phase and frequency locking. In this example, the mode with the longest lifetime will be the antisymmetric one. This, at sufficiently long times, implies that the LSPs of both NPs will end oscillating in anti-phase.

In general, systems with dynamical phase transitions are expected to have large differences in the imaginary parts of the eigenfrequencies, as in the case of coupled piano strings or in the example shown in Fig. \ref{FigureA1}. However, phase and frequency locking is not an exclusive phenomenom of this situation.
For the particular case of metallic nanoparticle arrays, the value of the damping terms needed to achieve the DPT described here are far from the realistic situation, at least for metallic NPs. Thus, the cases discussed in the main section of the article correspond to systems that do not present a DPT.

\end{document}